\documentstyle[aps,prc,epsfig]{revtex}
\newcommand{\beq}{\begin{equation}}
\newcommand{\eeq}{\end{equation}}
\newcommand{\bea}{\begin{eqnarray}}
\newcommand{\eea}{\end{eqnarray}}
\newcommand{\bce}{\begin{center}}
\newcommand{\ece}{\end{center}}
\newcommand{\eg}{{\it e.g.}}
\newcommand{\ie}{{\it i.e.}}
\newcommand{\etal}{{\it et al.}}
\def\lsim{\mathrel{\rlap{\lower4pt\hbox{\hskip1pt$\sim$}}
    \raise1pt\hbox{$<$}}}         
\def\gsim{\mathrel{\rlap{\lower4pt\hbox{\hskip1pt$\sim$}}
    \raise1pt\hbox{$>$}}}         
\begin{document}
\tightenlines
\title
{Thermal Dilepton Radiation at Intermediate Masses at the CERN-SpS}  
 
\author
{Ralf Rapp and Edward Shuryak}
 
\address
{Department of Physics and Astronomy, State University of New York, 
    Stony Brook, NY 11794-3800, U.S.A.}

\date{\today} 

\maketitle
 
\begin{abstract}
We investigate the significance of thermal dilepton radiation in the
intermediate-mass region in heavy-ion reactions at CERN-SpS energies.
Within  a thermal fireball model for the space-time evolution,  
the radiation from hot matter is found to dominate 
over hard 'background' processes (Drell-Yan and open charm)
up to invariant masses of about 2~GeV, with a rather moderate fraction
emerging from early stages with temperatures $T\simeq 175-200$~MeV 
associated with deconfined matter. 
Further including a schematic acceptance for the NA50 experiment we 
find good agreement with  the observed enhancement in the 
region 1.5~GeV~$<M_{\mu\mu}<$~3~GeV.
In particular, there is no need to invoke any 
anomalous open charm enhancement. 
\end{abstract}


\section{Introduction}
The identification of experimental signals from a possibly formed
Quark-Gluon Plasma (QGP) constitutes one of the prime goals in the 
(ultra-) relativistic heavy-ion program. 
Among the earliest suggestions~\cite{Shu80}
for QGP signatures has been the appearance 
of an enhanced radiation of dileptons (and/or photons) in the 
{\em intermediate}-mass region (IMR), 1.5~GeV~$<M_{ll}<$~3~GeV, 
\ie, well above the 
light vector mesons  $\rho$, $\omega$ and $\phi$ but below   
the $J/\Psi$ resonances. The importance of this region resides on the
facts that, on the one hand, low-energy hadronic processes (such as $\pi\pi$
annihilation or Dalitz decays) are sufficiently suppressed, and,  on the
other hand, hard processes (in particular Drell-Yan annihilation), 
which prevail in the high-mass region $M\ge 4$~GeV, increase rather slowly 
towards smaller $M$ and thus may be over-shined
by thermal radiation. 

At the CERN-SpS, intermediate-mass dilepton spectra have been measured
in the dimuon mode by the NA38/NA50~\cite{Scomp96,Bord99} 
and  HELIOS-3~\cite{HELIOS3} collaborations
(the dielectron data from the CERES experiment~\cite{CERES} lack  
sufficient statistics in this region). In central collisions
of heavy nuclei both experiments found factors of 2--3 enhancement over the 
extrapolation of known sources from proton-induced collisions,  
given by primordial Drell-Yan annihilation~\cite{DrYa} as well as 
semileptonic decays of associatedly produced charmed ($D$, $\bar D$) mesons.  

Several suggestions to explain this excess have been 
made, \eg, an enhanced production of $c\bar c$ pairs. However, 
such an increase is not easily justified theoretically~\cite{LMW95}, and
would also be difficult to reconcile with the current understanding
of the observed $J/\Psi$ {\em suppression}.  
Another possibility has been pursued by Lin and Wang~\cite{LW98} 
who showed that $D$-meson rescattering in hot/dense matter 
(without invoking anomalous enhancement) might generate a transverse momentum 
broadening which can enrich the $\mu^+\mu^-$ phase 
space covered by the NA50 experiment. The resulting increase is, however, 
much too small to explain the data~\cite{Bord99}.  
Spieles \etal~\cite{Spiel98} have investigated the  role of 
{\em secondary} Drell-Yan processes in hadronic rescatterings within the 
UrQMD transport model  and found that at an invariant mass of, 
\eg,  $M_{\mu\mu}=1.5$~GeV this contribution may constitute
up to $\sim$~30~\% of the primordial Drell-Yan yield. This is again far from 
being able to account for the experimental findings. 
Finally, Li and Gale analyzed the IMR of the HELIOS-3 data using 
a transport model incorporating dilepton production through 
secondary hadronic annihilation processes~\cite{LG98}. They found that 
the enhancement in S+W collisions 
can indeed be explained; unfortunately, the HELIOS-3
data quickly run out of statistics for $M_{\mu\mu}\gsim 2$~GeV.         
Since the transport framework employed was entirely based 
on hadronic degrees of freedom, a possible Quark-Gluon Plasma 
formation was not addressed explicitly. 

Another important issue is the consistency with dilepton measurements 
in the low-mass region (LMR). The CERES/NA45~\cite{CERES} and 
HELIOS-3~\cite{HELIOS3} collaborations
have reported a significant excess over final-state hadron decays which
can indeed be explained by thermal radiation once medium effects
in the vector meson spectral functions (especially for the $\rho$) are 
accounted for~\cite{RWX97,RW99} (see Ref.~\cite{RW00} for a recent review). 
In this letter we would like to investigate whether thermal production 
in the 1.5--3~GeV region, folded over the same fireball evolution
as employed for the LMR in Ref.~\cite{RW99}, can simultaneously  
describe the NA50 data. In addition, by allowing for a possible 
QGP formation, we study the significance of dileptons
produced through quark-antiquark annihilation in the early  phases
of central Pb(158~AGeV)+Pb collisions.  

\section{Dilepton Sources at Intermediate Masses} 
In the absence of any medium (or rescattering) effects, dilepton
production in hadronic collisions arises from either initial hard processes
or the decay of produced mesons. At invariant masses beyond 1.5~GeV, the 
contribution from light hadron decays (containing $u$, $d$ and $s$ quarks) 
sharply drops off either due their insufficient mass (as, \eg, for 
$\rho$, $\omega$ and $\phi$ mesons) or because of their increasing
total widths resulting in a small electromagnetic branching ratio
(as, \eg, for the $\rho'$ resonances). For masses between 1.5~GeV~$<M<$~3~GeV
the most important decay contribution therefore arises from the 
associated decay of charmed mesons $D$, $\bar D$ originating 
from  primordial production of a $c\bar c$ pair (the thermal abundance 
of $D$-mesons is negligible at SpS energies).  
The other significant 'background' source in the IMR is the well-known 
Drell-Yan (DY) process~\cite{DrYa}. 
The latter entirely saturates the yield at high 
masses beyond $\sim$5~GeV and can thus be exploited for normalization 
purposes as has been verified by the NA50 collaboration~\cite{na50-97}. 
To be able to  compare our calculations to the
NA50 data we will adopt the same strategy here 
(in addition, the shape of the DY-yield after applying 
acceptance cuts can be checked versus experimental simulations using 
event generators). 
In a central collision of two equal nuclei with mass number $A$ the 
number of produced Drell-Yan pairs per invariant mass and
rapidity interval can be obtained  as
\beq
\frac{dN_{DY}^{AA}}{dM dy}(b=0)=\frac{3}{4\pi R_0^2} A^{4/3}
\frac{d\sigma_{DY}^{NN}}{dM dy}
\label{dyaa}
\eeq
in terms of the (standard) elementary Drell-Yan cross section in a 
nucleon-nucleon collision, 
\beq
\frac{d\sigma_{DY}^{NN}}{dM dy}=K \frac{8\pi\alpha}{9sM} 
\sum\limits_{q=u,d,s} e_q^2 \left[ q(x_1) \bar q(x_2) + 
\bar q(x_1) q(x_2) \right]  \ . 
\label{dynn}
\eeq
Here, $q(x_{1,2})$ ($\bar q(x_{1,2})$) denote the (anti-) quark distribution 
functions within a pair of colliding nucleons from the two incoming 
nuclei (neglecting nuclear effects). Their arguments are related to the 
center-of-mass rapidity $y$ and invariant mass of the lepton 
pair as  $x_{1,2}=x e^{\pm y}$ with $x=M/\sqrt{s}$, where $s$ denotes 
the total $cms$ energy of the nucleon-nucleon collision. For the parton
distribution functions we employ the  recent leading order 
parameterization from Ref.~\cite{GRV95} (=GRV-94 LO), 
which necessitates a $K$-factor of 
1.5~\cite{Spiel98} (accounting for higher order effects) to comply  
with $p$-$A$ Drell-Yan data. As has been further pointed out in 
Ref.~\cite{Spiel98}, the explicit inclusion of isospin asymmetric
sea-quark distributions in the GRV-94 LO set entails substantially smaller 
isospin corrections for nuclear collisions; \eg, for Pb+Pb the correction
is as small as 3~\%, which will be neglected in the following. 
In Eq.~(\ref{dyaa}), the root-mean squared radius 
parameter $R_0\simeq1.05$~fm arises from a folding over a Gaussian
thickness function~\cite{Wong84}. 
For (slightly) non-central collisions as considered below we simply
replace $A$ in Eq.~(\ref{dyaa}) by the number of participants. 
  
In nuclear collisions additional so-called 'thermal' dileptons are 
radiated through 
interactions  during the expansion/cooling of the hadronic 
fireball. In thermal equilibrium, the dilepton production rate depends
on the  temperature and (baryon-) density of the matter and is governed
by the electromagnetic current-current correlator $\Pi_{\rm em}$. Whereas 
at low masses medium modifications of the light vector mesons 
(especially the $\rho$) lead to a strong reshaping
of the correlator~\cite{RW99}, the situation simplifies 
in the intermediate-mass region: already in vacuum the cross section for 
the reverse process of $e^+e^-\to hadrons$ is well-described by lowest-order
perturbative annihilation into $q\bar q$ pairs; at the correspondingly 
probed distance scales of $d\le 1/(1.5{\rm GeV})\simeq 0.13$~fm, 
additional medium effects are expected to be no longer relevant. In fact,
as has been shown in Ref.~\cite{LG98}, a rate calculation involving a large set
of hadronic reactions fitted to reproduce exclusive  $e^+e^-\to hadrons$ cross
sections yields an equivalent dilepton production rate (within the experimental
uncertainties), \ie, the hadronic and partonic description of $\Pi_{\rm em}$
are 'dual' for $M\ge 1.5$~GeV. We will thus employ the simple partonic 
rate
\bea
\frac{d^8N_{\mu\mu}}{d^4xd^4q} &=& \frac{\alpha^2}{4\pi^4} f^B(q_0;T)
\sum\limits_{q=u,d,s} (e_q)^2
\nonumber\\
           &=& \frac{\alpha^2}{4\pi^4} \ f^B(q_0;T) \ \frac{2}{3}
\label{rate}
\eea  
and note that in the IMR  both temperature and density corrections, being
of order $O(T/M)$ and $O(\mu_q/M)$,  are irrelevant under the 
conditions probed at the SpS.  
To obtain the total yield from thermal radiation, the rate (\ref{rate})
has to be integrated over the space-time history of a given collision
system. Here we employ a simplified thermal fireball description including, 
however, the most important features of a full hydrodynamic simulation
in a transparent way. As delineated in Ref.~\cite{RW99} we first 
construct a trajectory in the baryon-chemical potential/temperature plane
based on baryon-number and entropy conservation. Using a resonance gas 
equation of state (EoS) including the 32 (16) lowest lying mesonic (baryonic) 
states with  $s/\varrho_B$=26 it is found to be consistent with the 
recently determined {\em chemical} freezeout point of
$(\mu_B,T)_{chem}=(270\pm 20,175\pm 10)$~MeV~\cite{pbm98}, which we also
assume to be the critical point separating the hadronic and
quark-gluon phase. In the hadronic evolution towards {\it thermal}
freezeout,  $(T,\mu_N)_{fo}=(115\pm 10,430\pm 30)$~MeV, one additionally
requires pion (kaon) number conservation to maintain the final pion-to-baryon 
ratio of 5:1 at SpS energies, entailing the build-up of finite
pion (kaon) chemical potentials reaching $\mu_\pi=60-80$~MeV 
($\mu_{K,\bar K}=100-130$~MeV).  For the initial stages we assume an 
ideal QGP EoS. A time scale is introduced by employing a cylindrical 
hydro-type volume expansion 
\beq
V_{FC}= 2 (z_0+v_z t+\frac{1}{2} a_z t^2) \ 
\pi \ (r_0 + \frac{1}{2} a_\perp t^2)^2 \ 
\label{volexp}
\eeq
with a forward and a backward going firecylinder~\cite{RW99} to accomodate
a realistic rapidity spread in the final particle distributions.  
$r_0$ corresponds to the initial transverse overlap of the two colliding
nuclei (determined by the impact parameter $b$), and  the parameters 
$v_z$, $a_z$ and $a_\perp$ are adjusted to the finally observed
 flow velocities. 
Given the total entropy of the reaction, the value of $z_0$ is equivalent
to a formation time (or initial temperature).    
Once the system  has cooled down to $T_c$, the fraction $f$ of matter
in the hadronic phase can be inferred from the standard entropy balance
\beq
f s_{HG}(T_c) + (1-f) s_{QGP}(T_c) = S_{tot}/V_{FC}(t) \ .  
\eeq 
The time-integrated dilepton yield from one collision event then
takes the form
\beq
\frac{dN_{\mu\mu}^{therm}}{dM}=\frac{2}{3} \frac{\alpha^2}{\pi^3} M
\int\limits_0^{t_{fo}} dt \ V_{FC}(t) \int\limits_0^\infty dq 
 \ \frac{q^2}{q_0} \ f^B(q_0;T(t),\mu_\pi(t)) \ {\rm Acc}(M,q,y) \ , 
\eeq
where the factor Acc$(M,q,y)$ simulates the experimental acceptance
that the produced muons undergo in the NA50 detector, see below. 
In the hadronic phase, the Bose factor $f^B(q_0;T(t),\mu_\pi(t))$ 
includes the enhancement due to finite $\mu_\pi$'s. Since in the IMR 
the vector correlator is dominated by $\pi a_1$ annihilation~\cite{LG98}, 
which is a  
4-pion type process, the $\mu_\pi=0$ value of the Bose factor is 
augmented by a factor $\sim \exp(4\mu_\pi/T)$.  

\begin{figure}
\vspace{-1.5cm}
\bce
\epsfig{file=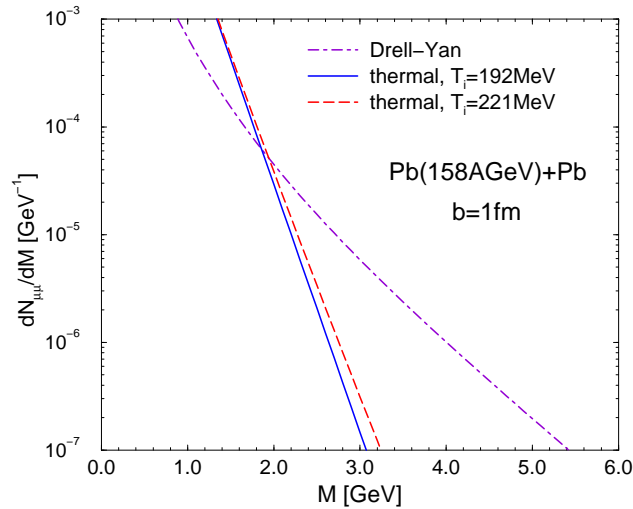,width=7.7cm,angle=-90}
\ece
\caption{Drell-Yan (dashed-dotted line) versus thermally produced
$\mu^+\mu^-$ pairs in central Pb(158AGeV)+Pb, integrated over all
rapidities and without any acceptance cuts.}    
\label{fig_nocuts}
\end{figure}

Before we come to the full results let us compare the relative
importance of Drell-Yan versus thermal pairs without any cuts. 
Fig.~\ref{fig_nocuts} shows the total dilepton yields from Drell-Yan
annihilation and the time-integrated thermal signal from a $b=1$~fm 
Pb(158~AGeV)+Pb collision. The latter has been calculated starting
from an initial temperature $T_i=192$~MeV (corresponding to $z_0=2.3$~fm 
in Eq.~(\ref{volexp})) cooling down to about 
$T_{fo}=113$~MeV within a total fireball lifetime of 
$\sim$~14~fm/c. As expected, the thermal curve (solid line) dominates towards
low invariant masses but drops below the DY component (dashed-dotted line) 
at $M\ge 1.8$~GeV due its much softer slope.   

To test the sensitivity with respect to the initial 
conditions, we have carried out a calculation with $z_0=1.5$~fm which
translates into an initial temperature $T_{i}=221$~MeV. With the same
fireball lifetime of 14~fm/c as before (leading to a practically identical 
freezeout temperature) the thermal spectrum looks rather similar to the  
$T_i=192$~MeV case up to $M\simeq 2$~GeV, indicating that the decrease 
of the rate at smaller temperatures (corresponding to later stages)
 is essentially compensated by the larger volumes. Only at 
masses beyond 2~GeV the higher initial temperature entails a visible increase 
of the thermal yield together with a slightly harder slope.    

\section{IMR Spectra in NA50}
In order to compare our calculations to the measurements of NA50
we have to account for the experimental acceptance cuts. In the IMR
the geometry of the spectrometer imposes kinematic restrictions which are 
approximately given by~\cite{Scomp99}    
\bea
0\le & y_{\mu\mu}^{cms} & \le 1 
\nonumber\\
-0.5 \le & \cos\Theta^+_{CS} & \le 0.5 \ , 
\label{cuts1}
\eea
where $y_{\mu\mu}^{cms}$ is the pair rapidity in the $cms$ of the colliding
nuclei and 
\beq
\cos\Theta^+_{CS}=\frac{p^+_z - \beta_z^{pair} E^+}{[(1-{\beta_z^{pair}}^2)
(M^2/4-m_\mu^2)]^{1/2}} \ .  
\eeq
Here, $\Theta^+_{CS}$ denotes the polar angle of the $\mu^+$ with respect
to the $z$-axis in the rest frame of the dimuon (the so-called Collins-Soper
frame) with $p^+_z$, $E^+$ and $\beta_z^{pair}=(p^+_z+p^-_z)/(E^++E^-)$
measured in the laboratory frame.  Furthermore, the single-muon tracks
in the lab-frame are selected according to
\beq
 0.037 \le  \Theta^\mu  \le 0.108 \ .  
\eeq
More difficult to asses are absorption effects on the single-muon tracks;
as a very rough characterization thereof we require the single-muon 
energies to satisfy~\cite{Scomp99}   
\beq
E_\mu  >    \left\{  
\begin{array}{ll}
 E_{cut} + 16000 (\Theta_\mu-0.065)^2 & , 0.037 \le \Theta_\mu \le 0.065 \\
 E_{cut} & , 0.065  \le \Theta_\mu \le 0.090   \\ 
 E_{cut} + 13000 (\Theta_\mu-0.090)^2 & , 0.090 \le \Theta_\mu \le 0.108 \ ,  
\end{array}  \right.
\label{cuts2}
\eeq
where $E_{cut}$ has been chosen as 8~GeV in Ref.~\cite{LW98}. 
Following Refs.~\cite{Scomp98,Bord99}, we have also incorporated  
a phenomenological
$q_t$-broadening  of the Drell-Yan spectra by supplying Eq.~(\ref{dyaa})
with a Gaussian factor $\exp(-q_t^2/2\sigma_{q_t}^2)/2\sigma_{q_t}^2$ 
to obtain $dN^{AA}_{DY}/dMdydq_t^2$ with  $\sigma_{q_t}=0.8-0.9$~GeV.  
Finally, we account for the finite experimental mass resolution by folding
the calculated spectra with a Gaussian, the effective $\sigma_M$ 
being estimated from the apparent width of the $J/\Psi$ peak.  

\begin{figure}
\vspace{-1.5cm}
\bce
\epsfig{file=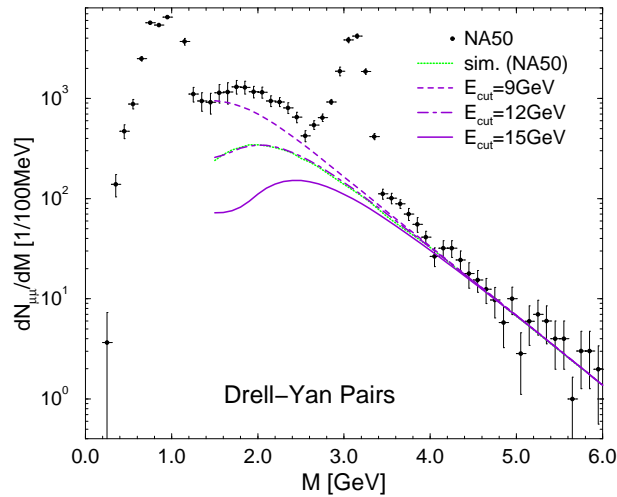,width=7.7cm,angle=-90}
\ece
\caption{Effects of the schematic cuts, Eqs.~(\protect\ref{cuts1}) and
(\protect\ref{cuts2}), on the invariant mass spectrum of Drell-Yan produced
pairs in central Pb(158AGeV)+Pb.
In comparison to a detector simulation of the NA50 collaboration
(dotted line; cf.~also Figure 1 in Ref.~\protect\cite{Scomp98} or
the Drell-Yan fits shown in Ref.~\protect\cite{Bord99})) the curve
calculated with $E_{cut}=12$~GeV seems to give the best  representation
of the experimental acceptance.}
\label{fig_dycuts}
\end{figure}
We have checked the schematic acceptance by comparing our results for 
the Drell-Yan
production with and without the cuts to a simulation of the NA50
collaboration using the PYTHIA event generator~\cite{Scomp98}.
The pertinent results are displayed in Fig.~\ref{fig_dycuts} using  
three different values for the single-muon energy threshold. The latter 
does not significantly affect the high-mass tail beyond 4~GeV which
has been used to fix the normalization in comparison to the 
NA50 data. From the behavior towards lower masses it seems that 
$E_{cut}\simeq 12$~GeV gives the closest reproduction of the simulation 
results from the NA50 collaboration~\cite{Scomp98,Bord99}.  
We will therefore use this value in the following also for the thermal 
radiation contribution, but note that this might not be 
an accurate representation of the actual experimental situation.  

To enable a more complete comparison with the NA50 data, the associated
decays of open charm mesons have to be included. We simply adopt 
this contribution as given in Ref.~\cite{Bord99} (without any anomalous 
enhancement factor).  

\begin{figure}
\vspace{-1.5cm}
\bce
\epsfig{file=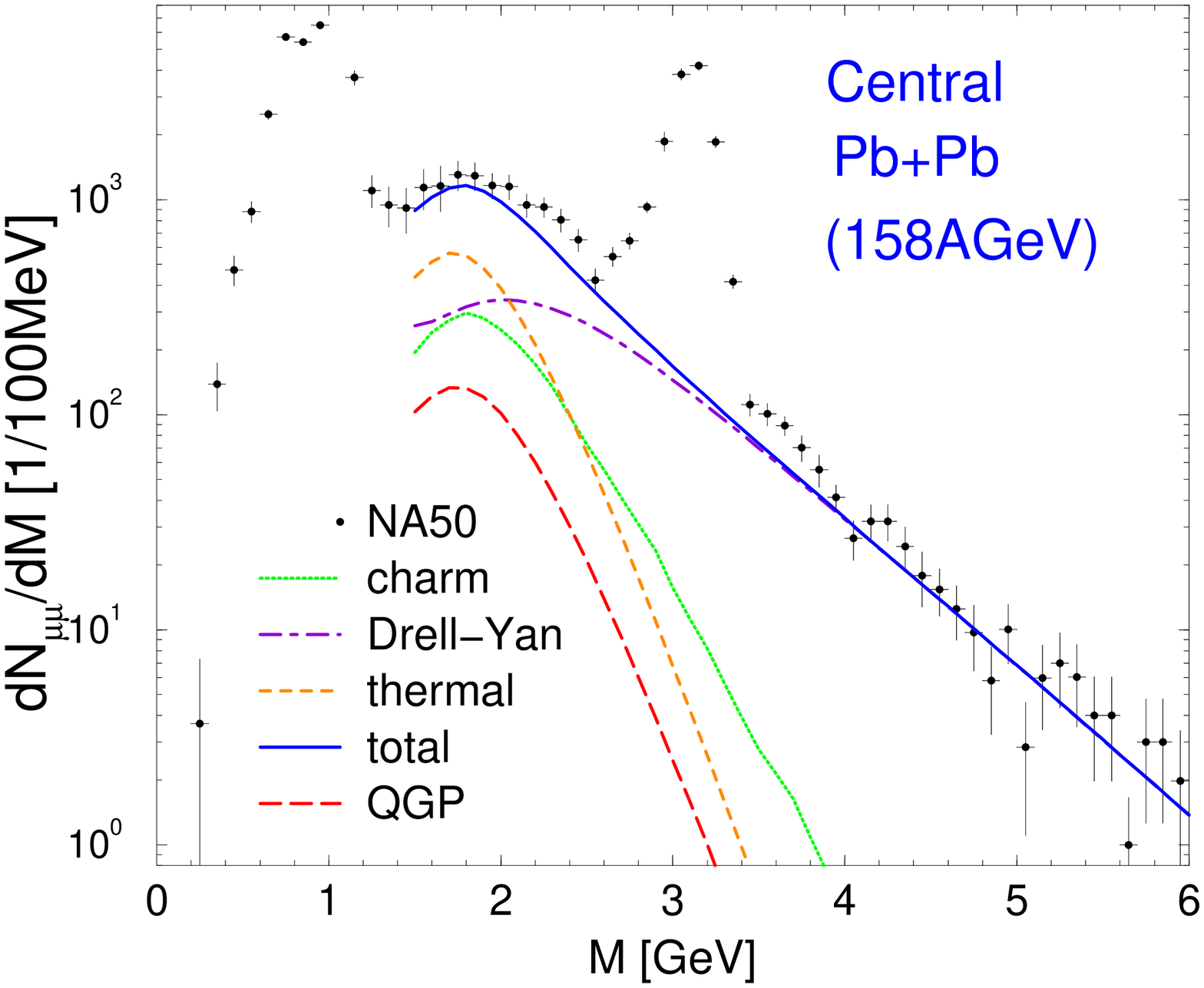,width=7.4cm,angle=-90}
\epsfig{file=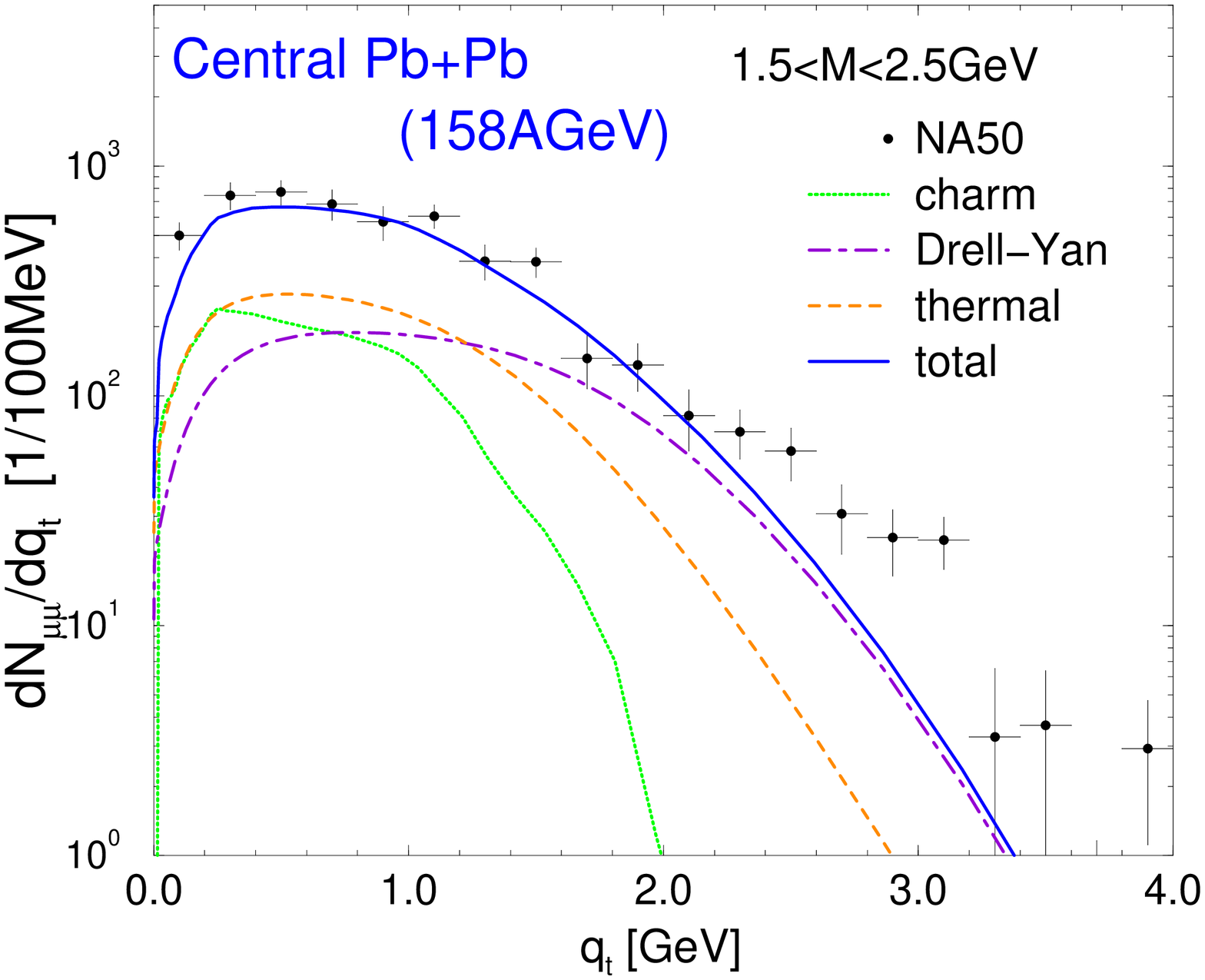,width=7.4cm,angle=-90}
\ece
\vspace{0.2cm}
\caption{IMR dilepton spectra in comparison to NA50 data from central
Pb(158~AGeV)+Pb collisions with $N_{part}\simeq 380$ (left panel: invariant
mass spectra, right panel: transverse momentum spectra). The short-dashed and
dashed-dotted curves are the calculated thermal and Drell-Yan yields,
respectively, using the approximate acceptance as given in the text.
The long-dashed curve in the left panel is the part of the
thermal contribution originating from the plasma phase. The dotted curves
represent the open charm contribution as obtained by the NA50 collaboration
in a PYTHIA simulation without anomalous enhancement~\protect\cite{Bord99}.
The full lines are the sum of thermal, open charm and Drell-Yan dileptons.}
\label{fig_dlna50}
\end{figure}

In Fig.~\ref{fig_dlna50} our final results are confronted with the NA50
data in central Pb+Pb collisions at 158~AGeV projectile energy. Given the
uncertainties in the simulation of the experimental acceptance, the excess 
observed in the IMR (1.5~GeV~$<M<$~2.5~GeV) can be well accounted for by  
thermal radiation when combined with Drell-Yan and open charm 
pairs (resulting in the full line in the left panel).
Also the shape and magnitude
of the transverse momentum spectra (right panel of Fig.~\ref{fig_dlna50})
are rather  well described once thermal pairs are included. 

As mentioned in the introduction an important issue in the  
IMR is the quantitative role of radiation from the QGP. With our standard
value for the initial temperature of $T_i=192$~MeV, the $q\bar q$
contribution to the thermal yield constitutes about 25~\% (slightly more
if a higher $T_i$ is assumed). This is significantly more than typical
estimates for the low-mass region, but still moderate  
owing to the rather small space-time volume occupied by a QGP at the
(full) SpS energies. Although perturbative $\alpha_s$-corrections 
might enhance the partonic production rate somewhat
the major signal will still originate from the hadronic phase.

One may further ask to what extent the presence of finite pion chemical 
potentials influences the thermal yield.  
When using $\mu_\pi=0$ throughout the latter is reduced by about 30--40~\%
(we should
stress, however, that such a calculation underestimates  
the measured pion multiplicity in central Pb(158~AGeV)+Pb by $\sim$~40--50~\%). 
Thus, similar to what has been found in the LMR~\cite{RW99}, chemical 
off-equilibrium effects (determined by imposing
pion number conservation) seem to be an important ingredient to understand 
dilepton production in the IMR.

\section{Conclusions}
In summary, we have evaluated dilepton spectra in central
Pb(158~AGeV)+Pb collisions as measured by the NA50 experiment. At intermediate
masses the total spectra are composed of 3 major components:
open charm decays, Drell-Yan annihilation and thermal radiation. 
Whereas the first one has been taken from NA50 event generator simulations,
the Drell-Yan part has been calculated explicitly to both determine the  
data normalization and check the approximate acceptance of the NA50 detector. 
The thermal contribution is based on a realistic fireball expansion in accord
with recent hadro-chemical analysis. Our main result is that the excess
observed by the NA50 collaboration in the mass region 1.5~GeV~$<M<$~2.5~GeV
can be explained by the thermal signal without invoking any anomalous
enhancement in the charm production. The importance of the thermal yield
quickly ceases beyond $M\simeq 2.5$~GeV, thus not affecting the interpretation 
of $J/\Psi$ suppression effects -- as opposed to scenarios invoking
open charm enhancement. The contribution from the QGP phase turned out to 
be around 20--30~\%. 

Since the low-mass dilepton spectra from CERES/NA45  
can be explained in the same framework (including medium effects in the
low-lying vector mesons) a consistent picture   of dilepton
production at the full CERN-SpS energy seems to emerge. Similar conclusions
have been reached in Ref.~\cite{GKP99} where a more schematic  
dynamical evolution has been employed.  

Let us finally address the question what can be done from the experimental
side to further test our results. Firstly, a direct measurement of $D$ mesons
will fix the open charm contribution to the dilepton spectra. Secondly, 
a dedicated (short) run of the NA50 experiment with reduced magnetic field
to focus on the IMR would be most valuable. The resulting gain in acceptance for
low-$q_t$ dileptons might quantitatively disentangle the thermal signal
from hot/dense matter, putting model predictions under scrutiny    
and increasing the sensitivity to the very early stages of the collision.

\vskip1cm
 
\centerline {\bf ACKNOWLEDGMENTS}
We are grateful for productive conversations with E. Scomparin, C. Gale,
B. K\"ampfer, C. Louren\c{c}o, H. Sorge and J. Wambach. 
One of us (RR) acknowledges support from the Alexander-von-Humboldt 
foundation as a Feodor-Lynen Fellow. 
This work is supported in part by the U.S. Department of Energy 
under Grant No. DE-FG02-88ER40388.

\end{document}